\documentclass[a4paper]{article}

\usepackage{INTERSPEECH2022}
\usepackage{threeparttable}
\usepackage{multicol}  
\usepackage{multirow}
\usepackage{hyperref} 
\usepackage{amsthm,amsmath,amssymb}
\usepackage{mathrsfs}
\usepackage{graphicx,subfigure}
\usepackage{makecell}
\usepackage{color}

\title{A Complementary Joint Training Approach Using Unpaired Speech and Text for Low-Resource Automatic Speech Recognition}
\name{Ye-Qian Du$^1$, Jie Zhang$^1$, Qiu-Shi Zhu$^1$, Li-Rong Dai$^1$, Ming-Hui Wu$^{2}$, Xin Fang$^{2}$, Zhou-Wang Yang$^1$}
\address{
  $^1$University of Science and Technology of China \\
  $^2$iFlytek Research}
\email{\{duyeqian,qszhu\}@mail.ustc.edu.cn, \{jzhang6,lrdai,yangzw\}@ustc.edu.cn}

\begin{document}
\setlength{\abovedisplayskip}{3.0pt}
\setlength{\belowdisplayskip}{3.0pt}
\maketitle


\begin{abstract}
  Unpaired data has shown to be beneficial for low-resource automatic speech recognition~(ASR), which can be involved in the design of hybrid models with multi-task training or language model dependent pre-training. In this work, we leverage unpaired data to train a general sequence-to-sequence model. Unpaired speech and text are used in the form of data pairs by generating the corresponding missing parts in prior to model training. Inspired by the complementarity of speech-PseudoLabel pair and SynthesizedAudio-text pair in both acoustic features and linguistic features, we propose a complementary joint training~(CJT) method that trains a model alternatively with two data pairs. Furthermore, label masking for pseudo-labels and gradient restriction for synthesized audio are proposed to further cope with the deviations from real data, termed as CJT++. Experimental results show that compared to speech-only training, the proposed basic CJT achieves great performance improvements on clean/other test sets, and the CJT++ re-training yields further performance enhancements. It is also apparent that the proposed method outperforms the wav2vec2.0 model with the same model size and beam size, particularly in extreme low-resource cases.
\end{abstract}
\noindent\textbf{Index Terms}: automatic speech recognition, low-resource, semi-supervised learning, speech synthesis, pseudo-label


\section{Introduction}

The end-to-end~(E2E) architecture remains the dominant paradigm for automatic speech recognition~(ASR). This single network structure allows for a simpler training process and joint optimization compared to conventional models, and it achieves impressive performances \cite{chan2016listen, dong2018speech, chiu2018state, zhang2017very}. Nevertheless, it requires a large amount of labeled data for training, which is rather expensive and time-consuming in terms of data collection, resulting in obstruction in the development of low-resource tasks. In contrast, speech-only and text-only data are broadly available. Thus, the focus of this work is on how to make use of unpaired data for low-resource ASR.

There has been extensive research on the utilization of unpaired data. For speech-only data, the common approach is unsupervised training that serves as a feature extractor for downstream ASR tasks \cite{van2018representation, Schneider2019, baevski2020wav2vec, hsu2021hubert}, or self-training with pseudo-labels following a typical teacher-student training scheme \cite{Xu2020, likhomanenko2020slimipl}. For text-only data, text is mainly used to train an external language model~(LM) for joint decoding \cite{Chorowski2017, Sriram2018, kannan2018analysis, shin2019effective, salazar2020masked}. In order to make use of both unpaired speech and text, many methods have recently been proposed, e.g., integration of a pre-trained acoustic model and LM \cite{yi2021efficiently, zheng2021wav, deng2021improving, chung2021splat}, cycle-consistency based dual-training \cite{tjandra2017listening, ren2019almost, karita2019semi, hori2019cycle}, and shared representation learning \cite{Renduchintala2018, drexler2018combining, karita2018semi, ao2021speecht5}, which rely on hybrid models with multi-task training and some of which become less effective in cases with a very limited amount of labeled data. The current mainstream methods that achieve state-of-the-art~(SOTA) results in low-resource ASR use unpaired speech and text for pre-training and training a LM for joint decoding, respectively \cite{baevski2020wav2vec, hsu2021hubert}, and adopt an additional iterative self-training \cite{xu2021self}. However, these methods require a large beam search space to fully exploit the capability of the LM, leading to a heavy computational cost for decoding or self-training.

\begin{figure}[t]\centering
\includegraphics[width=0.9\linewidth]{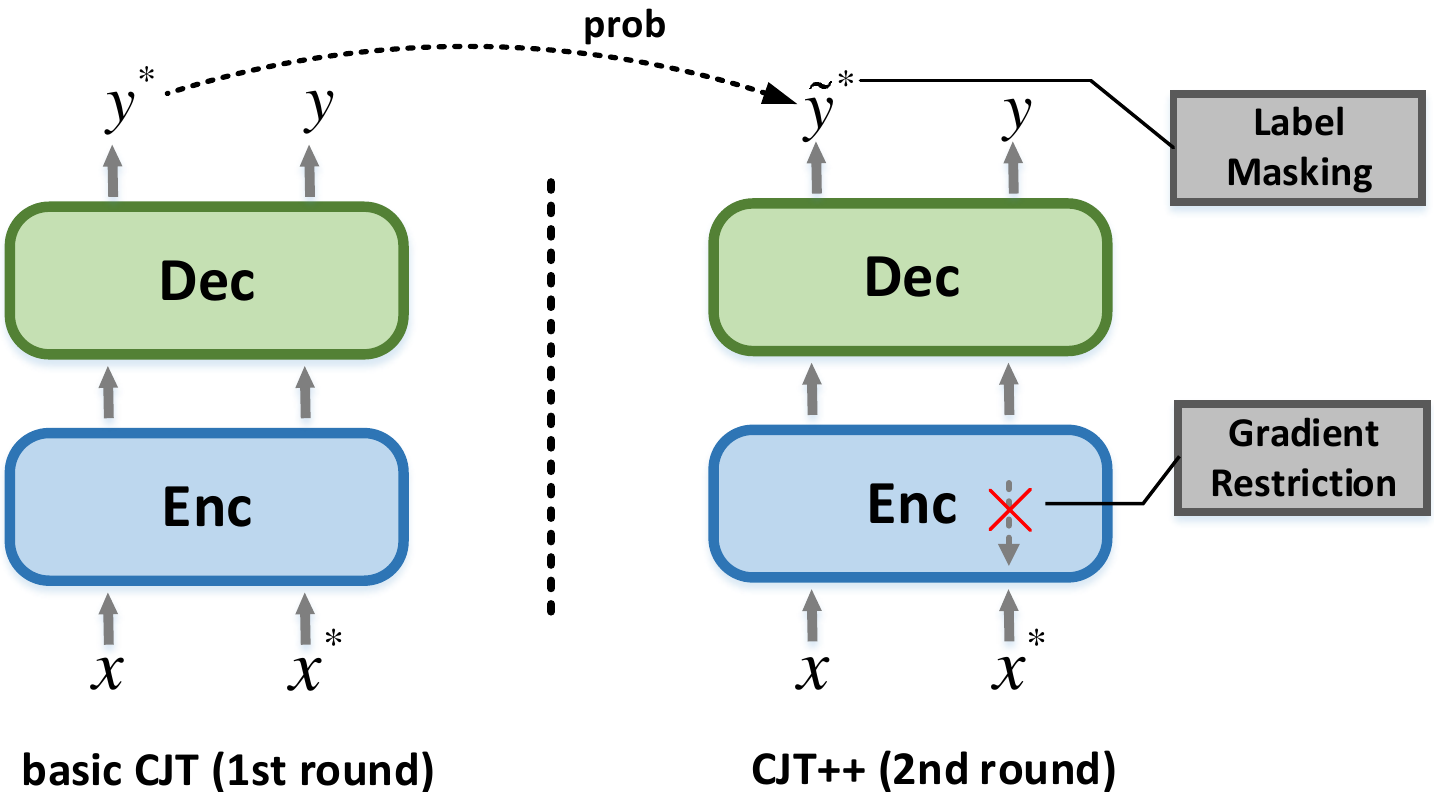}
\vspace{-0.2cm}
\caption{The diagram of the proposed CJT method, where $(x, y^*)$ is the speech-PseL pair that is generated by the ASR model fine-tuned on labeled data, and $(x^*, y)$ is the SynA-text pair generated by a TTS model.}
\vspace{-0.5cm}
\label{fig:pipline}
\end{figure}

In order to leverage unpaired speech and text to train a general sequence-to-sequence~(Seq2Seq) model, partial pre-training \cite{fan2019unsupervised, gao2021pre} can be applied, while it was shown that the inconsistency between pre-training and fine-tuning might limit the model performance.
To avoid such problem, in this work, we instead train the ASR model using sample pairs, which are generated by pseudo-labeling and Text-to-Speech~(TTS) synthesis prior to model training. As in low-resource scenarios, the generated data often largely differs from real data, so a single utilization of speech-PseudoLabel~(speech-PseL) pairs or SynthesizedAudio-text~(SynA-text) pairs could seriously mislead the model training. Due to the fact that these two kinds of data pairs are complementary in terms of both input acoustic features and output linguistic features, we propose to alternatively train the model on both data pairs, and we refer to this method as complementary joint training~(CJT). This CJT method is employed as the first round training and is thus called basic CJT. Based on the analysis of basic CJT, two strategies are proposed for further performance enhancement. Specifically, for pseudo-labels we mask the tokens with a low first-round confidence, and for synthesized audio we proportionally block the gradient back propagation to lower layers to better fit real audio. These two strategies are involved in the second round training, referred to as CJT++.

The proposed CJT method is validated via experiments on the LibriSpeech dataset \cite{panayotov2015librispeech} and a Transformer \cite{mohamed2019transformers} network with limited computational resources. Experimental results on the 10min labeled data show that the basic CJT reduces the word error rate~(WER) by around 35\%/21\% on clean/other sets compared to speech-only training, and the CJT++ re-training further reduces the WER by around 28\%/13\% , i.e., an overall 53\%/31\% reduction. It is also shown that on three low-resource data splits, the proposed method decreases the WER by 55\%/41\%/28\% on average on 10min/1h/10h labeled data compared to the wav2vec 2.0 model under the same modest model size and beam size. 

\vspace{-0.2cm}
\section{Complementary Joint Training}

\begin{figure}[t]\centering
\includegraphics[width=0.7\linewidth]{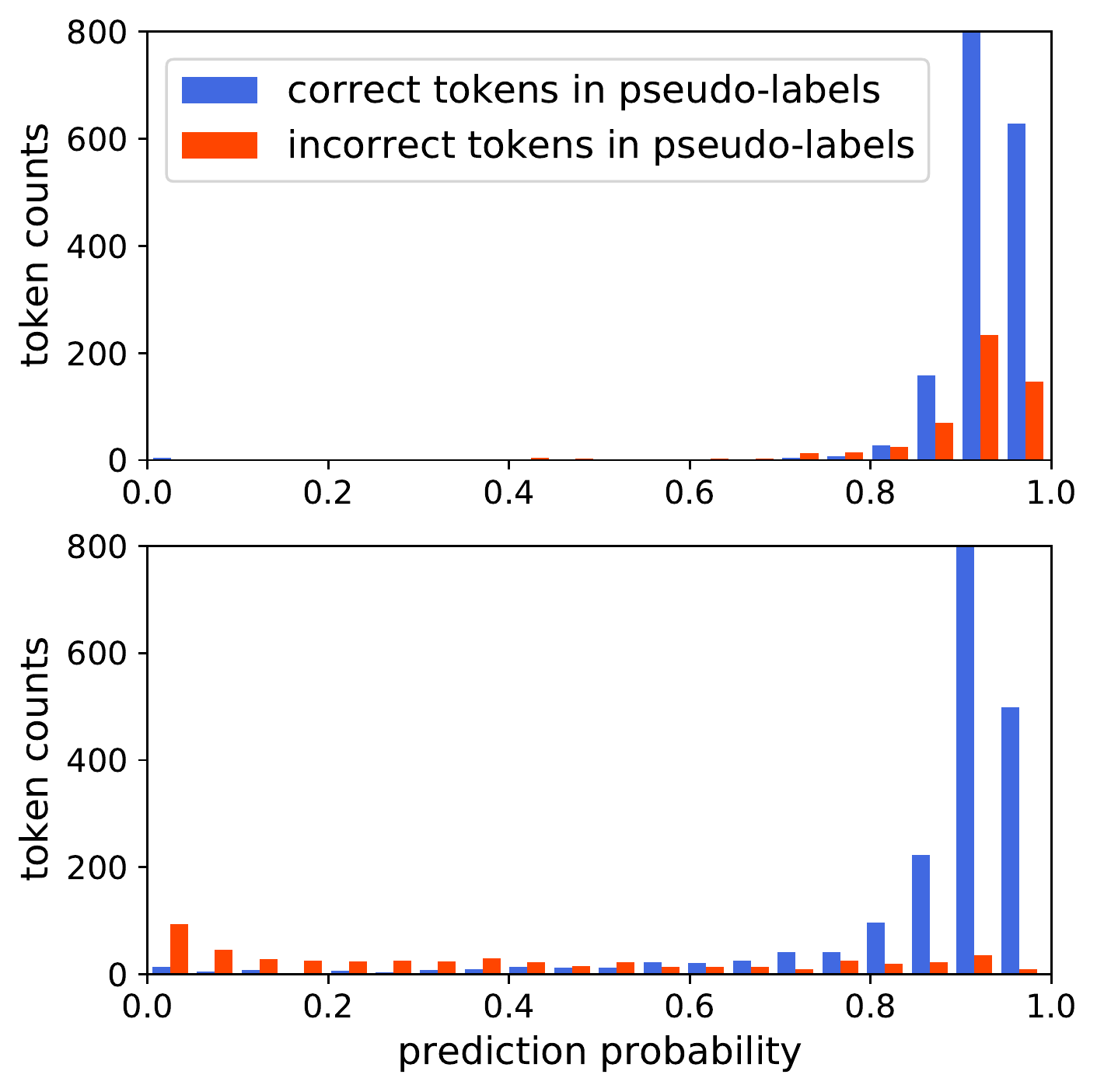}
\vspace{-0.2cm}
\caption{The predicted probabilities of correct and incorrect tokens. We use the model trained on speech-PseL data (upper) and the model jointly trained on both speech-PseL and SynA-text data (lower) to predict the probabilities of 4K pseudo-label tokens generated by 10min labeled data.}
\vspace{-0.5cm}
\label{fig:labelmask}
\end{figure}

The training process of the proposed CJT method is shown in Figure~\ref{fig:pipline}. After data preparation, the model is first jointly trained (basic CJT), as described in Section~\ref{sec:method1}, and then re-trained with label masking and gradient restriction (CJT++), as proposed in Section~\ref{sec:method2} after an empirical analysis.

\subsection{Basic complementary joint training}\label{sec:method1}

For the basic CJT, the abundant unpaired speech and text are used for training, and the small amount of paired data is only used for data preparation. Let the paired speech-text be denoted as $\mathcal{D}_p=\{(\mathbf{x}^{(i)}, \mathbf{y}^{(i)})\}^N_{i=1}$, the unpaired speech and unpaired text as $\mathcal{D}^s_u = \{\mathbf{x}^{(i)}\}^{N^s}_{i=1}$ and $\mathcal{D}^t_u = \{\mathbf{y}^{(i)}\}^{N^t}_{i=1}$, respectively. For an unpaired speech sample $\mathbf{x} \in \mathcal{D}^s_u$, we generate the corresponding pseudo-label $\mathbf{y}^*=\rm{ASR}^{(\mathcal{D}_p)}(\mathbf{x})$ by using the ASR model fine-tuned on the paired data $\mathcal{D}_p$. The set of speech-PseL pairs is denoted as ${\mathcal{D}^s_u}^*=\{(\mathbf{x}^{(i)},{\mathbf{y}^*}^{(i)} )\}^{N^s}_{i=1}$. For an unpaired text sample $\mathbf{y} \in \mathcal{D}^t_u$, we synthesize the corresponding audio $\mathbf{x}^*=\rm{TTS}(\mathbf{y})$ with a TTS model. The set of SynA-text pairs is denoted as ${\mathcal{D}^t_u}^*=\{({\mathbf{x}^*}^{(i)},{\mathbf{y}}^{(i)} )\}^{N^t}_{i=1}$. 

The CJT model is alternatively updated on ${\mathcal{D}^s_u}^*$ and ${\mathcal{D}^t_u}^*$, where the joint training target is given by
\begin{equation}
\mathcal{L} = \mathcal{L}_s + \lambda \mathcal{L}_t,
\end{equation}
where $\lambda$ is a balancing parameter, and $\mathcal{L}_s$ and $\mathcal{L}_t$ are losses of speech-PseL pairs and SynA-text pairs, respectively, given by
\begin{align}
\mathcal{L}_s &= -\mathbb{E}_{(\mathbf{x}, \mathbf{y}^*)\in {\mathcal{D}^s_u}^*} \log P(\mathbf{y}^*|\mathbf{x}),\\
\mathcal{L}_t &= -\mathbb{E}_{(\mathbf{x}^*, \mathbf{y})\in {\mathcal{D}^t_u}^*} \log P(\mathbf{y}|\mathbf{x}^*) .
\end{align}

\begin{figure}[t]\centering
\includegraphics[width=0.7\linewidth]{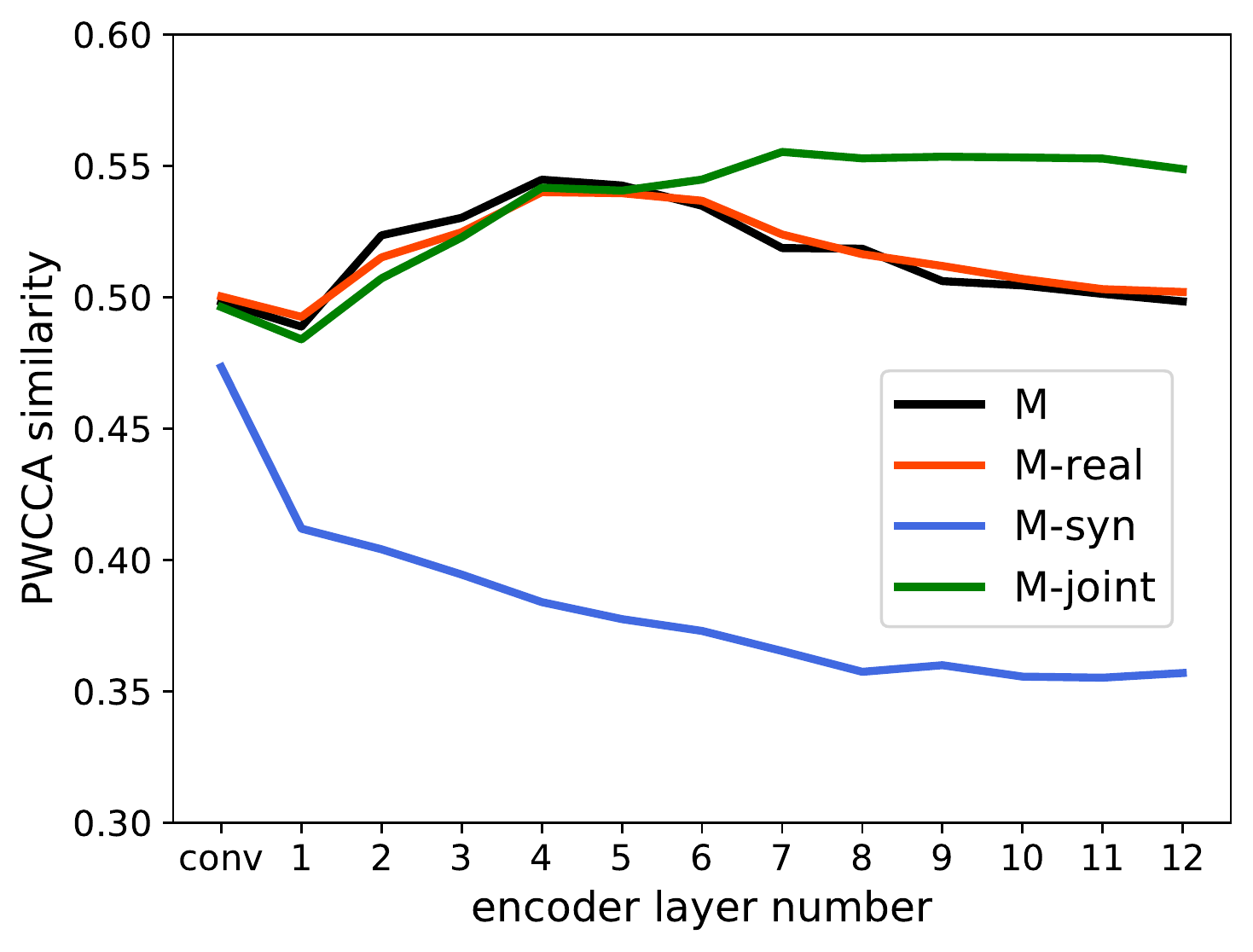}
\vspace{-0.2cm}
\caption{PWCCA similarity at encoder layers. We use 100K input frames from $dev\_clean$ real audio as the consistent input and extract the features of the model trained on (1) 100h paired data (M), (2) 100h speech-PseL data (M-real), (3) 860h SynA-text data (M-syn), (4) 100h speech-PseL data and 860h SynA-text data (M-joint). Then, we calculate the PWCCA with that of the standard model trained on 960h paired data.}
\vspace{-0.3cm}
\label{fig:pwcca}
\end{figure}

\subsection{Analysis and two enhancement strategies}\label{sec:method2}

In order to reveal the complementary properties of speech-PseL and SynA-text pairs, on one hand, we show the probabilities of the correctly and incorrectly predicted tokens in Figure~\ref{fig:labelmask}. It is clear that training on single pseudo-label data shows similar probability distribution between correct and incorrect tokens; while a joint training with real text largely reduces the probabilities of incorrect tokens. On the other hand, we show the PWCCA \cite{morcos2018insights} as a measure of similarity to compare representations at encoder layers with the classic supervised model in Figure~\ref{fig:pwcca}. This reveals that training on TTS synthesized audio causes an obvious deviation; while training on both synthesized audio and real audio allows the model to obtain many more similar features to the ground-truth at lower layers. Furthermore, there is even stronger linguistic modeling ability at higher layers. Altogether, it can be concluded that speech-PseL and SynA-text pairs are complementary nature in terms of linguistic token prediction and acoustic feature learning. Based on these observations, we propose two enhancement strategies for the basic CJT training as follows.

\subsubsection{Label masking for pseudo-labels}\label{sec:method2-1}

Training on pseudo-labels could overfit the model to errors in pseudo-labels, which would mislead the final recognition, and the additional text enables the model to be more discriminative in detecting incorrect tokens (e.g., see Figure~\ref{fig:labelmask}). We therefore take the prediction probability as a reference for identifying incorrect target tokens and propose a label masking strategy that masks the tokens with lower first-round prediction probability. The potential benefit of it is twofold: preventing overfitting to incorrect labels and an enhancing context modeling due to an absence of historical information.

For a pseudo-label sequence $\mathbf{y}^*=(y_1^*, y_2^* \cdots y_T^*)$, where $T$ is the length of the target sequence, we generate a binary mask sequence and accordingly replace some of the tokens with $<$PAD$>$. Let the masked target sequence be denoted as $\tilde{\mathbf{y}}^*$ and the set of mask indexes as $\mathcal{M}$, respectively. The loss of speech-PseL pairs in the second-round training becomes
\begin{equation}
\mathcal{L}_s'=-\mathbb{E}_{(\mathbf{x}, \mathbf{y}^*)\in {\mathcal{D}^s_u}}^* \sum_{t\in \{1,2 \cdots T\} \backslash \mathcal{M}}  \log P(\tilde{y}^*_t | \mathbf{x}, \tilde{\mathbf{y}}^*_{1:t-1} )
\end{equation}

Given the predicted probability $p$ of a token from the first-round prediction, three masking approaches are considered: 1) confidence-driven masking (\emph{conf}), where the token is randomly masked with the probability of $k*(1-p)$ with $k$ denoting a multiplier; 2) threshold-based masking (\emph{thres}), where the token is masked if $p \le P_\text{thres}$ with $P_\text{thres}$ denoting the threshold determined by the percentile of probabilities; 3) random masking (\emph{rand}), where the token is randomly masked with a fixed probability. The overall masking probability is empirically set as a multiple of the pseudo-label error rate.

\subsubsection{Gradient restriction for synthesized audio}\label{sec:method2-2}

Due to the fact that the TTS synthesized audio exhibits smaller variations than real audio, the utilization of synthesized data might degrade the ASR performance on real speech; however, after joint training with real audio, the model can largely recover the ability of acoustic feature extraction (e.g., see Figure~\ref{fig:pwcca}). 
Although synthesized audio causes only small deviations, we propose a gradient restriction training strategy to further reduce feature mismatches. 

In this strategy, we randomly block gradient propagation to the shallow layers when training on synthesized audio at a certain probability so that the model can better fit real audio when extracting acoustic features. In this work, the first four layers are regarded as the shallow layers.


\section{Performance Evaluation}

\subsection{Experimental setup}

All experiments are performed on the LibriSpeech corpus \cite{panayotov2015librispeech}. The unpaired speech originates from the LibriSpeech training data, which contains 960 hours of speech with transcriptions removed. The unpaired text comes from the standard pre-processed LibriSpeech LM corpus without overlapping transcripts, which contains about 80 times the amount of text in the audio transcriptions. Three Libri-light \cite{kahn2020libri} limited resource training subsets are used for the paired data, including train-10h (10 hours), train-1h (1 hour), and train-10min (10 minutes). Results are evaluated on dev-clean/other and test-clean/other sets. 

For the ASR modeling, we use 80-dimensional log-Mel filterbank features. The modeling units in our experiments are 5000 word pieces.
We choose the convolutional Transformer architecture \cite{mohamed2019transformers} as the backbone. This model (71M) is composed of an encoder that contains 2 2-D convolutional blocks \cite{mohamed2019transformers} followed by 12 Transformer blocks \cite{vaswani2017attention}, and a decoder that contains 4 1-D convolutional blocks \cite{mohamed2019transformers} followed by 6 Transformer blocks \cite{vaswani2017attention}. For each Transformer block, the attention dimension is 512 with 8 attention heads, and the inner dimension between layers is 2048. Besides, the self-attention is augmented with relative positional embedding \cite{shaw2018self}. 
The overall batch size is around 160. We use the Adam algorithm \cite{kingma2014adam} for optimization with a peak learning rate of 5e-4. A tri-stage learning rate scheduler \cite{park2019specaugment} is applied wherein the learning rate is linearly warmed up, kept constant, and then exponentially decayed for 10\%, 40\% and 50\% of the updates, respectively. Models on 100h/960h speech data are trained for 80K/250K updates. In case the model is jointly trained with text data, it trains for double updates. For the second-round training, we train for half the time of the first-round training. For regularization, we use a single dropout rate of 0.15 \cite{srivastava2014dropout} across all Transformer blocks and 0.1 label smoothing \cite{szegedy2016rethinking}. We also apply SpecAugment \cite{park2019specaugment} for data augmentation. The final model used for evaluation is calculated by averaging the last 10 checkpoints.

For language modeling, a Transformer-based LM consisting of 6 decoder blocks is trained on the LibriSpeech LM corpus. It trains for 800K updates under almost the same training conditions as in the ASR model. It shares the same 5000 word pieces as output tokens for shallow fusion \cite{Chorowski2017}. Finally, the LM has word-pieces-level perplexity of 32 on the dev-clean set. For decoding, the LM weight for shallow fusion is set to 0.4, with a beam size of 20.

\interfootnotelinepenalty=10000
For the preparation of pseudo-labels, we follow the well-designed pre-training model wav2vec 2.0 \cite{baevski2020wav2vec} \footnote{\url{https://github.com/pytorch/fairseq/tree/main/examples/wav2vec}} to train a teacher model for pseudo-labeling. The 960h pre-trained wav2vec 2.0 BASE model \footnote{\url{https://dl.fbaipublicfiles.com/fairseq/wav2vec/wav2vec_small.pt}} is loaded and fine-tuned on three labeled data splits with CTC loss \cite{graves2006connectionist}. Then, we pseudo-label the 960h unpaired speech by the fine-tuned models combined with a loaded word-level Transformer LM \footnote{\url{https://dl.fbaipublicfiles.com/wav2letter/sota/2019/lm/lm_librispeech_word_transformer.pt}}, using a medium beam size of 20. The WERs of pseudo-labels of 10h, 1h, 10min labeled data are 4.96\%, 7.96\% and 17.10\%, respectively.

For the preparation of synthesized audio, we use a ready-made TTS engine with 3 speakers \footnote{Refer to \url{http://ttsvoice.iflysec.com/}} for convenience, as the paired data that are less than 10 hours or even 10 minutes are hard to train a robust TTS model. Only 1/10 of the LM corpus is randomly selected for audio synthesis.

Notice that all experiments are implemented in the fairseq framework \cite{ott2019fairseq}. Training and decoding hyper-parameters are barely tuned for  better possible performance.

\subsection{Results}


\begin{table}[t]
  \caption{Performance of the joint training with different amounts of unpaired data and various updating ratios (1:$\mathbf{\lambda}$) using 10min of labeled data, where 200h of unpaired speech consists of the train-clean-100 set and 100h of the rest clean data. All results are obtained by ASR-only greedy decoding.}
  \vspace{-0.2cm}
  \label{tab:exp1}
  \centering
  \begin{threeparttable} 
  \resizebox{0.47\textwidth}{25mm}{
    \begin{tabular}{ccccccc}
    \toprule[1pt]
    \multicolumn{2}{c}{\textbf{Unpaired Data}} &  \multirow{2}{*}{1:$\mathbf{\lambda}$} & \multicolumn{2}{c}{\textbf{Dev WER}} & \multicolumn{2}{c}{\textbf{Test WER}}  \cr
    \cmidrule(lr){1-2} \cmidrule(lr){4-5} \cmidrule(lr){6-7} 
    speech-PseL & SynA-text & & clean & other & clean & other \cr
    \midrule[1pt]
    100h & - & - & 25.18 & 40.20 & 25.72 & 41.72  \cr
    200h & - & - & 18.88 & 31.19 & 19.36 & 31.69 \cr
    960h & - & - & 15.47 & 23.37 & 16.03 & 23.30  \cr
    \midrule
    - & 100h & - & 92.99 & 95.47 & 94.16 & 95.23 \cr
    - & 860h & - & 92.27 & 96.01 & 94.56 & 96.01 \cr
    \midrule
    100h & 100h & 1:1 & 18.68 & 33.66 & 18.77 & 34.89 \cr
    100h & 860h & 1:1 & 17.29 & 32.14 & 17.51 & 33.26 \cr
    100h & 860h & 1:3 & 16.31 & 32.16 & 16.72 & 32.61 \cr
    100h & 860h & 1:5 & 15.96 & 32.24 & 16.59 & 33.09 \cr
    \midrule[1pt]
    \multicolumn{2}{c}{100h (speech-text)} & - & 13.90 & 30.46 & 13.91 & 31.30  \cr
    \bottomrule[1pt]
    \end{tabular} 
    }
  \end{threeparttable}
  \vspace{-0.3cm}
\end{table}

First of all, in order to validate the effectiveness of the basic CJT method with different amounts of data and updating ratios, we show the ASR performance in terms of WER for the proposed CJT, the models trained on a single type of data and the oracle model trained with ground-truth transcriptions in Table~\ref{tab:exp1}. We see that the WERs of 100h speech-PseL + 100h SynA-text are comparable to that of 200h speech-PseL (although slightly higher on noisy(other) set). If the amount of SynA-text is increased to 860h, the performance is further improved, e.g., even very close to the WER of 960h speech-PseL on the clean set, but obviously worse on the noisy(other) set. Among the three updating ratios, 1:3 performs better and if it raises up to 1:5, the results get better on the clean set but worse on the noisy(other) set. Overall, joint training with additional text achieves a relative WER reduction by around 35\%/21\% on clean/other sets. The reason for the smaller improvement on the noisy(other) set might be that there are fewer variations in the synthesized audio when using text data.



\begin{table}[t]
  \vspace{-0.2cm}
  \caption{Performance of the proposed second-round training strategies with various settings, following the training in Table~\ref{tab:exp1}. All results are obtained by ASR-only greedy decoding.}
  \vspace{-0.2cm}
  \label{tab:exp2}
  \centering
  \begin{threeparttable} 
  \resizebox{0.45\textwidth}{26mm}{
    \begin{tabular}{lcccc}
    \toprule[1pt]
    \multirow{2}{*}{\textbf{Method}} & \multicolumn{2}{c}{\textbf{Dev WER}} & \multicolumn{2}{c}{\textbf{Test WER}}  \cr
    \cmidrule(lr){2-3} \cmidrule(lr){4-5} 
     & clean & other & clean & other \cr
    \midrule[1pt]
    basic CJT & 16.31 & 32.16 & 16.72 & 32.61 \cr
    \midrule
    \quad + PseLM-\emph{rand}(p=0.4) & 15.57 & 31.15 & 16.02 & 32.22 \cr
    \quad + PseLM-\emph{conf}(p=0.4) & 12.93 & 28.92 & 13.17 & 29.73  \cr
    \quad + PseLM-\emph{thres}(p=0.16) & 12.85 & 28.84 & 13.44 & 29.57 \cr
    \quad + PseLM-\emph{thres}(p=0.4) & 11.84 & 28.29 & 12.05 & 29.52 \cr
    \quad + PseLM-\emph{thres}(p=0.8) & 11.67 & 30.08 & 12.08 & 31.49 \cr
    \midrule
    \quad + SynGR-\emph{all} & 16.46 & 30.71 & 16.76 & 31.54 \cr
    \quad + SynGR-\emph{shallow} & 16.27 & 30.24 & 16.57 & 31.23  \cr
    \midrule
    \makecell[l]{ \quad + PseLM-\emph{thres}(p=0.4) \\ \qquad \& SynGR-\emph{shallow}} & 11.74 & 27.74 & 12.01 & 28.83 \cr
    \bottomrule[1pt]
    \end{tabular} }
  \end{threeparttable}
  \vspace{-0.3cm}
\end{table}

Second, we show the performance of the proposed two second-round training strategies in comparison with the basic CJT method in Table~\ref{tab:exp2}. 
In label masking for pseudo-labels (PseLM), the three masking methods described in Section~\ref{sec:method2-1} are compared at a certain overall masking probability $p$. It is clear that threshold-based masking performs better and achieves obvious improvement over the basic CJT method, especially on the clean set. Compared to random masking, this performance gain is mainly due to the superior discrimination of target errors brought by the first-round training. It is interesting that in the basic CJT+PseLM-\emph{thres} method, increasing the masking probability from 0.16, the error rate of 100h pseudo-labels, to 0.4 results in better performance, indicating that over masking is more effective. The gain brought about by random masking or over masking reveals that label masking strategy also improves context modeling capability. While further increasing the probability from 0.4 to 0.8 causes performance degradation on noisy(other) sets.

In gradient restriction for synthesized audio (SynGR), we block the gradient propagation to 1-4 encoder layers (\emph{shallow}) and widen to all encoder layers (\emph{all}) for comparison at a probability of 0.7. It can be observed that gradient restriction in shallow layers yields slightly better performance, which is consistent with the PWCCA behavior in Figure~\ref{fig:pwcca}. SynGR only introduces a small improvement on noisy(other) sets, since as expected there is only a small gap that we can compensate (e.g., see Figure~\ref{fig:pwcca}).
Furthermore, applying both PseLM and SynGR in the proposed CJT method results in the proposed CJT++ approach. Compared to the basic CJT method, it is clear that CJT++ re-training with additional strategies relatively reduces the WER by around 28\%/13\% on clean/other sets, and the overall relative WER reduction over the speech-only results is around 53\%/31\% on clean/other sets.



\begin{table}[t]
  \vspace{-0.2cm}
  \caption{A comparison of WERs on Librispeech with the wav2vec 2.0 BASE model with the same beam size of 20, where ``*" stands for self-implementation. The models are trained on 3 Libri-light low-resource data splits, using the 960h untranscribed LibriSpeech data and the LM corpus as unpaired data.}
  \vspace{-0.2cm}
  \label{tab:exp3}
  \centering
  \begin{threeparttable} 
  \resizebox{0.48\textwidth}{49mm}{
    \begin{tabular}{lccccc}
    \toprule[1pt]
    \multirow{2}{*}{\textbf{Method}} & \multirow{2}{*}{\textbf{LM}} & \multicolumn{2}{c}{\textbf{Dev WER}} &\multicolumn{2}{c}{\textbf{Test WER}}  \cr
    \cmidrule(lr){3-4} \cmidrule(lr){5-6} 
    & & clean & other & clean & other \cr
    \midrule[1pt]
    \textbf{10min paired}& &  &  &  &  \cr
    \midrule
    wav2vec 2.0 \cite{baevski2020wav2vec} & - & 46.1 & 51.5 & 46.9 & 50.9 \cr
    wav2vec 2.0 * & - & 47.43 & 53.56 & 48.31 & 53.21 \cr
    wav2vec 2.0 * & Transf.& 17.62 & 26.00 & 17.80 & 25.46 \cr
    \midrule
    basic CJT  & \multirow{3}{*}{Transf.} & 13.22 & 19.54 & 13.20 & 19.75 \cr
    \quad + CJT++ & & 10.02 & 16.32 & 10.14 & 16.61 \cr
    \quad \quad + LM& & 6.90 & 12.51 & 7.36 & 12.96 \cr
    \midrule
    \midrule
    \textbf{1h paired}& & &  &  &  \cr
    \midrule
    wav2vec 2.0 \cite{baevski2020wav2vec} & - &  24.1 & 29.6 & 24.5 & 29.7 \cr
    wav2vec 2.0 * & - & 18.21  & 25.67 & 18.86 & 26.29 \cr
    wav2vec 2.0 * &Transf. & 7.67 & 14.42 & 7.77 & 14.80 \cr
    \midrule
    basic CJT & \multirow{3}{*}{Transf.} & 6.44 & 12.36 & 6.45 & 12.78 \cr
    \quad + CJT++ & & 5.68 & 11.58 & 5.66 & 12.19 \cr
    \quad \quad + LM& & 4.25 & 8.74 & 4.25 & 9.44 \cr
    \midrule
    \midrule
    \textbf{10h paired}& & &  &  &  \cr
    \midrule
    wav2vec 2.0 \cite{baevski2020wav2vec} & - & 10.9 & 17.4 & 11.1 & 17.6 \cr
    wav2vec 2.0 * & - & 9.51 & 17.00 & 9.76 & 17.30 \cr
    wav2vec 2.0 * & Transf. &  4.64 & 10.59 & 4.62 & 10.68 \cr
    \midrule
    basic CJT & \multirow{3}{*}{Transf.} & 4.40 & 10.20 & 4.44 & 10.64 \cr
    \quad + CJT++ & & 4.13 & 9.77 & 4.22 & 10.34 \cr
    \quad \quad + LM& & 3.03 & 7.66 & 3.42 & 8.30 \cr
    \bottomrule[1pt]
    \end{tabular}}
  \end{threeparttable}
  \vspace{-0.3cm}
\end{table}

Finally, we validate the proposed CJT method on three low-resource labeled data splits in comparison with the wav2vec 2.0 BASE model in Table~\ref{tab:exp3}. The proposed training method decoding with LM by shallow fusion reaches WERs of 7.36\%/12.96\%,  4.25\%/9.44\%, and 3.42\%/8.30\% on 10min, 1h, 10h labeled data of test clean/other, respectively, using a moderate model size and beam size. Note that only 1/10 of the LM corpus is used in the ASR training by CJT, thus LM fusion can still bring a quite obvious improvement. For fair comparison, we implement the wav2vec 2.0 BASE model with the same beam search size as a reference. It is clear that CJT performs better, particularly in the extreme low-resource case of 10min paired data.



\section{Conclusions}

In this paper, we proposed a CJT-based semi-supervised approach for low-resource ASR, which includes a first-round basic CJT and a second-round CJT++ re-training with two strategies, i.e., label masking and gradient restriction. It was shown that the joint training of two generated data pairs is complementary and compatible in both analytic and experimental fashions. Re-training with label masking and gradient restriction can further enhance the effectiveness of CJT. In order to study the robustness of the proposed CJT method, in the future, we will make efforts to optimize the beam size for pseudo-labeling, as well as the in-domain TTS model for speech synthesis in comparison with prior SOTA results, e.g., in \cite{baevski2020wav2vec, xu2021self, hsu2021hubert}.

\section{Acknowledgements}


We would like to thank IFLYTEK CO. LTD. for providing computational resources and a TTS engine. This work is supported by Anhui Center for Applied Mathematics, the Strategic Priority Research Program of Chinese Academy of Sciences (No. XDC 08010100), the NSF of China (No. 11871447), and the National Natural Science Foundation of China (No. 62101523).

\bibliographystyle{IEEEtran}

\bibliography{mybib}

\begin{thebibliography}{10}
\providecommand{\url}[1]{#1}
\csname url@samestyle\endcsname
\providecommand{\newblock}{\relax}
\providecommand{\bibinfo}[2]{#2}
\providecommand{\BIBentrySTDinterwordspacing}{\spaceskip=0pt\relax}
\providecommand{\BIBentryALTinterwordstretchfactor}{4}
\providecommand{\BIBentryALTinterwordspacing}{\spaceskip=\fontdimen2\font plus
\BIBentryALTinterwordstretchfactor\fontdimen3\font minus
  \fontdimen4\font\relax}
\providecommand{\BIBforeignlanguage}[2]{{%
\expandafter\ifx\csname l@#1\endcsname\relax
\typeout{** WARNING: IEEEtran.bst: No hyphenation pattern has been}%
\typeout{** loaded for the language `#1'. Using the pattern for}%
\typeout{** the default language instead.}%
\else
\language=\csname l@#1\endcsname
\fi
#2}}
\providecommand{\BIBdecl}{\relax}
\BIBdecl

\bibitem{chan2016listen}
W.~Chan, N.~Jaitly, Q.~Le, and O.~Vinyals, ``Listen, attend and spell: A neural
  network for large vocabulary conversational speech recognition,'' in
  \emph{ICASSP}, 2016.

\bibitem{dong2018speech}
L.~Dong, S.~Xu, and B.~Xu, ``Speech-transformer: A no-recurrence
  sequence-to-sequence model for speech recognition,'' in \emph{ICASSP}, 2018.

\bibitem{chiu2018state}
C.-C. Chiu, T.~N. Sainath, Y.~Wu, R.~Prabhavalkar, P.~Nguyen \emph{et~al.},
  ``State-of-the-art speech recognition with sequence-to-sequence models,'' in
  \emph{ICASSP}, 2018.

\bibitem{zhang2017very}
Y.~Zhang, W.~Chan, and N.~Jaitly, ``Very deep convolutional networks for
  end-to-end speech recognition,'' in \emph{ICASSP}, 2017.

\bibitem{van2018representation}
A.~Van~den Oord, Y.~Li, and O.~Vinyals, ``Representation learning with
  contrastive predictive coding,'' \emph{arXiv}, 2018.

\bibitem{Schneider2019}
S.~Schneider, A.~Baevski, R.~Collobert, and M.~Auli, ``wav2vec: Unsupervised
  pre-training for speech recognition,'' in \emph{Interspeech}, 2019.

\bibitem{baevski2020wav2vec}
A.~Baevski, Y.~Zhou, A.~Mohamed, and M.~Auli, ``wav2vec 2.0: A framework for
  self-supervised learning of speech representations,'' \emph{Advances in
  Neural Information Processing Systems}, 2020.

\bibitem{hsu2021hubert}
W.-N. Hsu, B.~Bolte, Y.-H.~H. Tsai, K.~Lakhotia, R.~Salakhutdinov, and
  A.~Mohamed, ``{HuBERT}: Self-supervised speech representation learning by
  masked prediction of hidden units,'' \emph{IEEE/ACM Transactions on Audio,
  Speech, and Language Processing}, 2021.

\bibitem{Xu2020}
Q.~Xu, T.~Likhomanenko, J.~Kahn, A.~Hannun, G.~Synnaeve, and R.~Collobert,
  ``Iterative pseudo-labeling for speech recognition,'' in \emph{Interspeech},
  2020.

\bibitem{likhomanenko2020slimipl}
T.~Likhomanenko, Q.~Xu, J.~Kahn, G.~Synnaeve, and R.~Collobert, ``{SLIMIPL}:
  Language-model-free iterative pseudo-labeling,'' \emph{arXiv}, 2020.

\bibitem{Chorowski2017}
J.~Chorowski and N.~Jaitly, ``Towards better decoding and language model
  integration in sequence to sequence models,'' in \emph{Interspeech}, 2017.

\bibitem{Sriram2018}
A.~Sriram, H.~Jun, S.~Satheesh, and A.~Coates, ``Cold fusion: Training seq2seq
  models together with language models,'' in \emph{Interspeech}, 2018.

\bibitem{kannan2018analysis}
A.~Kannan, Y.~Wu, P.~Nguyen, T.~N. Sainath, Z.~Chen, and R.~Prabhavalkar, ``An
  analysis of incorporating an external language model into a
  sequence-to-sequence model,'' in \emph{ICASSP}, 2018.

\bibitem{shin2019effective}
J.~Shin, Y.~Lee, and K.~Jung, ``Effective sentence scoring method using {BERT}
  for speech recognition,'' in \emph{Proceedings of The Eleventh Asian
  Conference on Machine Learning}, 2019.

\bibitem{salazar2020masked}
J.~Salazar, D.~Liang, T.~Q. Nguyen, and K.~Kirchhoff, ``Masked language model
  scoring,'' in \emph{Annual Meeting of the Association for Computational
  Linguistics}, 2020.

\bibitem{yi2021efficiently}
C.~Yi, S.~Zhou, and B.~Xu, ``Efficiently fusing pretrained acoustic and
  linguistic encoders for low-resource speech recognition,'' \emph{IEEE Signal
  Processing Letters}, 2021.

\bibitem{zheng2021wav}
G.~Zheng, Y.~Xiao, K.~Gong, P.~Zhou, X.~Liang, and L.~Lin, ``Wav-{BERT}:
  Cooperative acoustic and linguistic representation learning for low-resource
  speech recognition,'' in \emph{Findings of the Association for Computational
  Linguistics: EMNLP}, 2021.

\bibitem{deng2021improving}
K.~Deng, S.~Cao, Y.~Zhang, and L.~Ma, ``Improving hybrid ctc/attention
  end-to-end speech recognition with pretrained acoustic and language model,''
  \emph{arXiv}, 2021.

\bibitem{chung2021splat}
Y.-A. Chung, C.~Zhu, and M.~Zeng, ``{SPLAT}: Speech-language joint pre-training
  for spoken language understanding,'' in \emph{NAACL-HLT}, 2021.

\bibitem{tjandra2017listening}
A.~Tjandra, S.~Sakti, and S.~Nakamura, ``Listening while speaking: Speech chain
  by deep learning,'' in \emph{IEEE Automatic Speech Recognition and
  Understanding Workshop (ASRU)}, 2017.

\bibitem{ren2019almost}
Y.~Ren, X.~Tan, T.~Qin, S.~Zhao, Z.~Zhao, and T.-Y. Liu, ``Almost unsupervised
  text to speech and automatic speech recognition,'' in \emph{International
  Conference on Machine Learning}, 2019.

\bibitem{karita2019semi}
S.~Karita, S.~Watanabe, T.~Iwata, M.~Delcroix, A.~Ogawa, and T.~Nakatani,
  ``Semi-supervised end-to-end speech recognition using text-to-speech and
  autoencoders,'' in \emph{ICASSP}, 2019.

\bibitem{hori2019cycle}
T.~Hori, R.~Astudillo, T.~Hayashi, Y.~Zhang, S.~Watanabe, and J.~Le~Roux,
  ``Cycle-consistency training for end-to-end speech recognition,'' in
  \emph{ICASSP}, 2019.

\bibitem{Renduchintala2018}
A.~Renduchintala, S.~Ding, M.~Wiesner, and S.~Watanabe, ``Multi-modal data
  augmentation for end-to-end {ASR},'' in \emph{Interspeech}, 2018.

\bibitem{drexler2018combining}
J.~Drexler and J.~Glass, ``Combining end-to-end and adversarial training for
  low-resource speech recognition,'' in \emph{IEEE Spoken Language Technology
  Workshop (SLT)}, 2018.

\bibitem{karita2018semi}
S.~Karita, S.~Watanabe, T.~Iwata, A.~Ogawa, and M.~Delcroix, ``Semi-supervised
  end-to-end speech recognition.'' in \emph{Interspeech}, 2018.

\bibitem{ao2021speecht5}
J.~Ao, R.~Wang, L.~Zhou, S.~Liu, S.~Ren, Y.~Wu, T.~Ko, Q.~Li, Y.~Zhang, Z.~Wei
  \emph{et~al.}, ``{SpeechT5}: Unified-modal encoder-decoder pre-training for
  spoken language processing,'' \emph{arXiv}, 2021.

\bibitem{xu2021self}
Q.~Xu, A.~Baevski, T.~Likhomanenko, P.~Tomasello, A.~Conneau \emph{et~al.},
  ``Self-training and pre-training are complementary for speech recognition,''
  in \emph{ICASSP}, 2021.

\bibitem{fan2019unsupervised}
Z.~Fan, S.~Zhou, and B.~Xu, ``Unsupervised pre-training for sequence to
  sequence speech recognition,'' \emph{arXiv}, 2019.

\bibitem{gao2021pre}
C.~Gao, G.~Cheng, R.~Yang, H.~Zhu, P.~Zhang, and Y.~Yan, ``Pre-training
  transformer decoder for end-to-end {ASR} model with unpaired text data,'' in
  \emph{ICASSP}, 2021.

\bibitem{panayotov2015librispeech}
V.~Panayotov, G.~Chen, D.~Povey, and S.~Khudanpur, ``Librispeech: An {ASR}
  corpus based on public domain audio books,'' in \emph{ICASSP}, 2015.

\bibitem{mohamed2019transformers}
A.~Mohamed, D.~Okhonko, and L.~Zettlemoyer, ``Transformers with convolutional
  context for {ASR},'' \emph{arXiv}, 2019.

\bibitem{morcos2018insights}
A.~Morcos, M.~Raghu, and S.~Bengio, ``Insights on representational similarity
  in neural networks with canonical correlation,'' \emph{Advances in Neural
  Information Processing Systems}, 2018.

\bibitem{kahn2020libri}
J.~Kahn, M.~Riviere, W.~Zheng, E.~Kharitonov, Q.~Xu \emph{et~al.},
  ``Libri-light: A benchmark for {ASR} with limited or no supervision,'' in
  \emph{ICASSP}, 2020.

\bibitem{vaswani2017attention}
A.~Vaswani, N.~Shazeer, N.~Parmar, J.~Uszkoreit, L.~Jones \emph{et~al.},
  ``Attention is all you need,'' \emph{Advances in Neural Information
  Processing Systems}, 2017.

\bibitem{shaw2018self}
P.~Shaw, J.~Uszkoreit, and A.~Vaswani, ``Self-attention with relative position
  representations,'' in \emph{NAACL-HLT}, 2018.

\bibitem{kingma2014adam}
D.~P. Kingma and J.~Ba, ``Adam: A method for stochastic optimization,''
  \emph{arXiv}, 2014.

\bibitem{park2019specaugment}
D.~S. Park, W.~Chan, Y.~Zhang, C.-C. Chiu, B.~Zoph, E.~D. Cubuk, and Q.~V. Le,
  ``Specaugment: A simple data augmentation method for automatic speech
  recognition,'' \emph{Interspeech}, 2019.

\bibitem{srivastava2014dropout}
N.~Srivastava, G.~Hinton, A.~Krizhevsky, I.~Sutskever, and R.~Salakhutdinov,
  ``Dropout: a simple way to prevent neural networks from overfitting,''
  \emph{The journal of machine learning research}, 2014.

\bibitem{szegedy2016rethinking}
C.~Szegedy, V.~Vanhoucke, S.~Ioffe, J.~Shlens, and Z.~Wojna, ``Rethinking the
  inception architecture for computer vision,'' in \emph{IEEE Conference on
  Computer Vision and Pattern Recognition (CVPR)}, 2016.

\bibitem{graves2006connectionist}
A.~Graves, S.~Fern{\'a}ndez, F.~Gomez, and J.~Schmidhuber, ``Connectionist
  temporal classification: labelling unsegmented sequence data with recurrent
  neural networks,'' in \emph{international conference on Machine learning},
  2006.

\bibitem{ott2019fairseq}
M.~Ott, S.~Edunov, A.~Baevski, A.~Fan, S.~Gross, N.~Ng, D.~Grangier, and
  M.~Auli, ``fairseq: A fast, extensible toolkit for sequence modeling,''
  \emph{NAACL}, 2019.

\end{thebibliography}

\end{document}